\documentstyle[prl,aps]{revtex}

\begin{document}

\draft
\title{On the low-temperature phase of the three-state antiferromagnetic
Potts model on the simple cubic lattice}
\author{Alessandro Pelizzola}
\address{Dipartimento di Fisica and Istituto Nazionale per la Fisica 
della Materia, c.\ Duca degli Abruzzi 24, I-10129 Torino, Italy}
\maketitle

\begin{abstract}
The three-state antiferromagnetic Potts model on the simple cubic lattice
is investigated using the cluster variation method in the cube and 
the star-cube approximations. The broken-sublattice-symmetry phase is found 
to be stable in the whole low-temperature region, contrary to previous 
results obtained using a modified cluster variation method. The tiny free 
energy difference between the broken-sublattice-symmetry and the 
permutationally-symmetric-sublattices phases is calculated in the two 
approximations and turns out to be smaller 
in the (more accurate) star-cube approximation 
than in the cube one. 
\end{abstract}

The three-state antiferromagnetic Potts model on the simple cubic lattice
has been intensively studied in the last two 
decades, because of its unusual and rather obscure low-temperature and 
critical properties
\cite{berkad,bagrja1,greban,bagrja2,ono,uesuon,waswko1,waswko2,kikoka,uenkas,roslap1,roslap2,gothas,kolsuz1,kolsuz2,kularo}. 
The model hamiltonian is very simple and can be
written as 
\begin{equation}
{\cal H} = J \sum_{\langle i j \rangle} \delta(s_i, s_j),
\label{ham}
\end{equation}
where $J>0$ is the interaction strength, the summation is over all nearest neighbour (NN) pairs, 
$\delta$ denotes Kronecker's delta, and the variables $s_i$ can take on 
three different values, say $\{1,2,3\}$. 

It is now commonly accepted that the model exhibits a continuous phase
transition at temperature $T_c = 1.23$ (in units of $J/k_B$, with $k_B$
Boltzmann's constant), belonging to the universality class of the
three-dimensional XY model, as suggested by the value of the critical
exponents recently calculated by means of extensive Monte Carlo
simulations \cite{gothas} and coherent-anomaly method \cite{kolsuz1}.
The nature of the ordered phase, however, is not yet as clear as the
critical behaviour.

In the past some authors suggested that the ordered phase should be
one with algebraically decaying correlations like a
Kosterlitz-Thouless phase \cite{berkad,ono,waswko1,waswko2}, or an
incompletely ordered phase \cite{uenkas}. Other studies, however,
indicated the existence of a long-range ordered phase called
broken-sublattice-symmetry (BSS) phase
\cite{bagrja1,greban,bagrja2,kikoka}. In this phase the simple cubic
lattice is split into two interpenetrating 
sublattices, say A and B, such that a site in A has all its nearest 
neighbours belonging to B and viceversa.
In terms of the site expectations
$p_{k,\alpha} = \langle \delta(s_i,k) \rangle, i \in \alpha$, with $k
= 1, 2, 3$ and $\alpha = A, B$, the BSS phase can be roughly
described, at very low temperatures, by $p_{1,A} = 1, p_{1,B} = 0$ and
$p_{k,A} = 0, p_{k,B} = 1/2$ for $k = 2, 3$ (it can be easily verified
that the phase is sixfold degenerate). To be more
precise, even at very low temperature, there is a small probability to
find $s_i = 2$ or 3 in sublattice A, too, because this causes no increase 
in energy when the neighboring sites are all in the state 3 or 2, respectively.
This implies that the long range order does not saturate even at zero
temperature, and that the zero temperature entropy per site is
slightly larger than the value $\frac{1}{2} \ln 2 \simeq 0.346574$ one
would predict naively. 

Furthermore, in a recent investigation of the Blume-Emery-Griffiths
model by Rosengren and Lapinskas \cite{roslap1}, based on a modified
cluster variation method, a new phase was found in an intermediate
temperature region ($0.78 < T < T_c$) between the BSS and the
disordered phase. This phase, of degeneracy twelve, is characterized
by the relation $p_{1,A} = p_{2,B} > p_{3,A} = p_{3,B} > p_{2,A} =
p_{1,B}$ between the site expectations, and has been called
permutationally-symmetric-sublattices (PSS) phase by the authors. The
result is qualitatively similar to that obtained in the Bethe
approximation (which, of course, predicts higher transition
temperatures), and have received a partial confirmation by a Monte
Carlo investigation by the same authors and Kundrotas \cite{kularo}.
On the other hand, a different Monte Carlo investigation by Kolesik
and Suzuki \cite{kolsuz2} suggests that an intermediate phase might
exist but cannot be the PSS phase predicted by Rosengren and
Lapinskas.  Kolesik and Suzuki have computed the free energies of the
BSS and PSS phases, finding indications that the BSS free energy is
lower in the low temperature region, up to a temperature which is
higher (close below 1.0) than the transition temperature proposed by
Rosengren and Lapinskas. For even higher temperatures, the free energy
difference between the BSS and PSS phases is indistinguishable from
zero, although the authors conclude that they cannot definitely rule
out the possibility that it vanishes only at the critical point.

In the present work we examine carefully the issue of the type of ordered 
phase(s) by means of two high order approximations of the cluster variation
method (CVM) \cite{kikuchi1,an,morita}. 
The CVM is a simple and powerful variational method for Ising-like models,
based on a cumulant expansion of the variational principle of statistical 
mechanics. The full (and generally unaffordable) variational principle
is based on the minimization of the functional
\begin{equation}
{\cal F}[\rho_\Lambda] = {\rm Tr}(\rho_\Lambda {\cal H} + k_B T \rho_\Lambda \ln
\rho_\Lambda),
\label{varprin}
\end{equation}
where $\Lambda$ is the lattice, $\rho_\Lambda$ the corresponding trial density
matrix and ${\rm Tr}$ stands for trace. In the CVM one
chooses a set of maximal clusters to take into account and considers the 
approximate variational functional
\begin{equation}
F[\{\rho_\alpha, \alpha \in M \}] = \sum_{\alpha \in M} 
{\rm Tr}(\rho_\alpha {\cal H}_\alpha + k_B T a_\alpha \rho_\alpha \ln
\rho_\alpha),
\label{varcvm}
\end{equation}
where $M$ is the set of the maximal clusters and all their subclusters, 
$\rho_\alpha$ the reduced trial density matrix for the cluster $\alpha$,
${\cal H}_\alpha$ the hamiltonian contribution associated to the cluster 
$\alpha$ (in the present case ${\cal H}_\alpha = 0$ if $\alpha$ is not a n.n. 
pair) and the $a_\alpha$'s are constant, geometry-dependent coefficients that can be easily  
obtained by solving a suitable set of linear equations \cite{an,morita}. 
The approximate functional $F$ has to be minimized with respect to the cluster
density matrices (which, for a classical model, are diagonal), with the 
constraints (normalization and compatibility, respectively)
\begin{equation}
{\rm Tr} \rho_\alpha = 1, \quad \alpha \in M \qquad {\rm and} \qquad
{\rm Tr}_{\alpha \setminus \beta} \rho_\alpha = \rho_\beta, \quad 
\beta \subset \alpha \in M.
\label{constr}
\end{equation}

In this work we shall use two different approximations in the CVM
scheme, the cube and the star-cube ones.
The cube approximation \cite{kikuchi1} is obtained by selecting as maximal clusters 
the cubic cells of the lattice, and the corresponding functional (free energy 
per site) is
\begin{equation}
f[\rho_8] = 3 {\rm Tr} (\rho_2 {\cal H}_2) + k_B T \left[
{\rm Tr}(\rho_8 \ln \rho_8) - 3 {\rm Tr}(\rho_4 \ln \rho_4) + 3 {\rm Tr}
(\rho_2 \ln \rho_2) - \frac{1}{2} {\rm Tr} (\rho_{1A} \ln \rho_{1A})
- \frac{1}{2} {\rm Tr} (\rho_{1B} \ln \rho_{1B}) \right],
\label{fcube}
\end{equation}
where $\rho_8$, $\rho_4$, $\rho_2$ and $\rho_{1 \alpha}$ denote the density 
matrices for the cube, square, n.n. pair and site (with a sublattice index 
$\alpha = A, B$) clusters, respectively. $f$ can be regarded as
a functional of $\rho_8$ only, since the compatibility constraints can be 
easily solved by {\em defining} the other density matrices as partial traces
of $\rho_8$. In principle, $\rho_8$ has $3^8 = 6561$ different diagonal 
elements, but symmetry considerations reduce this number to 495, which (apart
from the normalization constraint, which can be easily dealt with) is the final 
number of independent variables for this approximation. Despite this large 
number of variables, the local minima corresponding to the various phases are 
easily obtained by means of the so-called natural iteration method
(NIM) \cite{kikuchi2}.

The star-cube approximation \cite{peliz1} goes a step further, including in the set of 
maximal clusters also the ``stars'' formed by a site and its six nearest 
neighbours. This choice seems particularly useful here, since in view of the above 
remarks on the very low temperature properties of the BSS phase it should be
particularly important to consider explicitly the local environment of a 
site. The resulting approximate functional has the form
\begin{eqnarray}
f[\rho_8,\rho_{7A},\rho_{7B}] = && 3 {\rm Tr} (\rho_2 {\cal H}_2) + k_B T \Bigg[
{\rm Tr}(\rho_8 \ln \rho_8) + \frac{1}{2} {\rm Tr} (\rho_{7A} \ln
\rho_{7A}) +  \frac{1}{2} {\rm Tr} (\rho_{7B} \ln \rho_{7B}) - 3 
{\rm Tr}(\rho_4 \ln \rho_4) \nonumber \\ 
&&  - 4 {\rm Tr}(\rho_{4^\prime A} \ln
\rho_{4^\prime A}) - 4 {\rm Tr}(\rho_{4^\prime B} \ln
\rho_{4^\prime B}) + 6 {\rm Tr}(\rho_{3A} \ln \rho_{3A}) 
+ 6 {\rm Tr}(\rho_{3B} \ln \rho_{3B}) - 3 {\rm Tr}
(\rho_2 \ln \rho_2) \Bigg],
\label{fstarcube}
\end{eqnarray}
where $\rho_{7\alpha}$, $\rho_{4^\prime \alpha}$ and $\rho_{3 \alpha}$
are the reduced density matrices (all with a sublattice index $\alpha$
referring to the central site) for the star cluster and its four- and
three-site subclusters obtained by taking the central site and three or
two of its nearest neighbors forming a solid angle or a plane angle,
respectively. 
The functional now depends on $\rho_8$, $\rho_{7A}$ and $\rho_{7B}$
(subcluster matrices being defined as partial traces, as above), but
the diagonal elements of these matrices are not all independent
because of the compatibility constraints. Taking into account the
lattice symmetries one finds 495 variables for the cube matrix, 168
variables for each star matrix, and two sets of 30 linear constraints.
A more detailed account of  our calculations will be given elsewhere
\cite{accomp}. 

It is important to notice here that the calculation 
which suggested the stability of the PSS phase \cite{roslap1,roslap2} 
was based on a modification 
of the cluster variation method which, although considering many-point 
clusters (up to the cube cluster, like our cube approximation), does
not take into account properly 
the $n$-point correlations with $n > 2$. In the modified CVM, in fact,
the density matrices are written as $\rho_\alpha = \exp(-H_\alpha^{(\rm
eff)})$, where $H_\alpha^{(\rm eff)}$ is an effective hamiltonian
which contains only one-site and two-site terms. This results in a
reduced number of variational parameters and in a violation of the
compatibility constraints in Eq.\ \ref{constr} (for a full description
of the method see Ref.\ 
\cite{zublap}; the authors of Refs.\ \cite{roslap1,roslap2} do not report
details of the method used, leaving them to Ref.\ \cite{roslap3},
which in turn refers to \cite{zublap}). 
The results for the ordered phases are then qualitatively similar to those 
obtained using the much simpler two-point (Bethe) approximation. The
present work can then be regarded as an improvement of that described
in \cite{roslap1} for two reasons: first, the CVM is used in its full
form, i.e. without the restriction inherent to the modified version
used in \cite{roslap1}; second, with the star-cube approximation we go
one step further in the cumulant expansion of the variational principle.

Let us now turn to a brief description of our results.
As far as very low temperatures are concerned, results from the cube
approximation are in perfect agreement with the modified CVM at the
cube level \cite{roslap1,roslap2}. In
particular we have $p_{1A} = 0.939067$, $p_{2A} = p_{3A} = 0.030467$ and 
$p_{1B} = 0.000117$, $p_{2B} = p_{3B} = 0.499942$, while the
entropy per site is $s = 0.366928$.
The star-cube approximation introduces only very small corrections,
giving $p_{1A} = 0.939041$, $p_{2A} = p_{3A} = 0.030480$ and 
$p_{1B} = 0.000055$, $p_{2B} = p_{3B} = 0.499972$, with an 
entropy per site is $s = 0.366941$.

Also the critical temperature does not change much with respect to the
modified CVM. In the cube approximation we obtain $T_c = 1.268$, which is just 0.5 \%
higher than the modified CVM results, a difference of the same sign
and order of magnitude of the one found in the simple Ising case \cite{zublap}. In
the star-cube approximation the critical temperature is
slightly lower, $T_c \simeq 1.263$. These results are to be
compared with the most recent Monte Carlo estimate $T_c = 1.23$ \cite{gothas}. 

These estimates can be further improved by employing 
the recently introduced cluster-variation--Pad\`e-approximants method (CVPAM) 
\cite{peliz2,peliz3}, from which one can
try to extract also estimates for the critical exponents.  Using the
star-cube results for the susceptibility for values of $w = {\rm
tanh}(1/T)$ up to $w_{\rm max} = 0.48$ we have obtained (with simple
Dlog Pad\`e approximants) a critical temperature $T_c \simeq 1.24$ and
a critical exponent $\gamma$ ranging from 1.30 to 1.33, in agreement
with the best recent estimates yielding $\gamma \simeq 1.31$
\cite{gothas}. These CVPAM results are still affected by relatively
large sistematic errors, and an accurate CVPAM analysis would require
both using larger maximal clusters and including corrections to
scaling \cite{peliz3}. Improving in this direction is beyond the scope
of the present paper, but our calculation of $\gamma$ can at least be
regarded as a check of the accuracy of the CVM results.

The central result of our paper, qualitatively different from the modified
CVM, is however that the free energy of the BSS phase is always lower
(although slightly) than that of the PSS phase. In Fig.\ \ref{fig} we have
plotted the free energy difference $\delta f$ vs.\ temperature as given by the
two approximations. It is important to notice
that the numerical errors involved in the estimation of the free
energy are several order of magnitudes smaller than $\delta f$.
From Fig.\ \ref{fig} one can see that $\delta f$
is very small in the cube approximation, and becomes
even smaller in the more accurate star-cube approximation. In both
approximations $\delta f$ vanishes only at the critical point, but in
the star-cube approximation it becomes negligibly small at a
temperature close below 1.0, like in the Monte Carlo simulations by
Kolesik and Suzuki \cite{kolsuz2}. Furthermore, the numerical values of 
$\delta f$ are of the same order of magnitude of those
reported by these authors. 


Although it is difficult to judge on the basis of our results which is
the exact ordered phase, some conclusions can certainly be drawn.
First of all, the results found by Rosengren and
Lapinskas with the modified CVM \cite{roslap1} are ruled out and 
attributed to the neglect of high-order correlations (remember that a
simple Bethe approximation yields the same qualitative results). This
means also that the phase diagram they proposed for the
Blume-Emery-Griffiths model \cite{roslap2} is incorrect and must be
reconsidered. On the other hand, the present 
CVM predictions are compatible with the Monte Carlo results by
Kolesik and Suzuki \cite{kolsuz2}. The Monte Carlo simulations
tends to favor the possibility that the free energy difference $\delta f$
vanishes below the critical point, giving rise to a new intermediate
phase. The CVM results seems to support this possibility, the free
energy difference $\delta f$ being strongly depressed, while going from
the cube to the star-cube approximation, in the same temperature
region. In order to make this explicit, we have plotted in Fig.\
\ref{fig-ratio} the ratio between the two values of $\delta f$
reported in Fig.\ \ref{fig}. It can be seen that this ratio increases
very quickly for temperatures between 0.9 and 1.0, and this could be a
signal that going to larger and larger maximal clusters the free
energy difference tends to zero. Both the Monte Carlo simulations and
the cluster variation method, however, leave open also the possibility
that the free energy difference vanishes only at the critical point,
and further investigations are therefore welcome.

\begin{figure}
\caption{Free energy difference between the PSS and the BSS phases, as
given by the cube (dashed line) and the star-cube approximation (solid
line)}
\label{fig}
\end{figure}

\begin{figure}
\caption{Ratio of the free energy differences given by the cube and
the star-cube approximations}
\label{fig-ratio}
\end{figure}

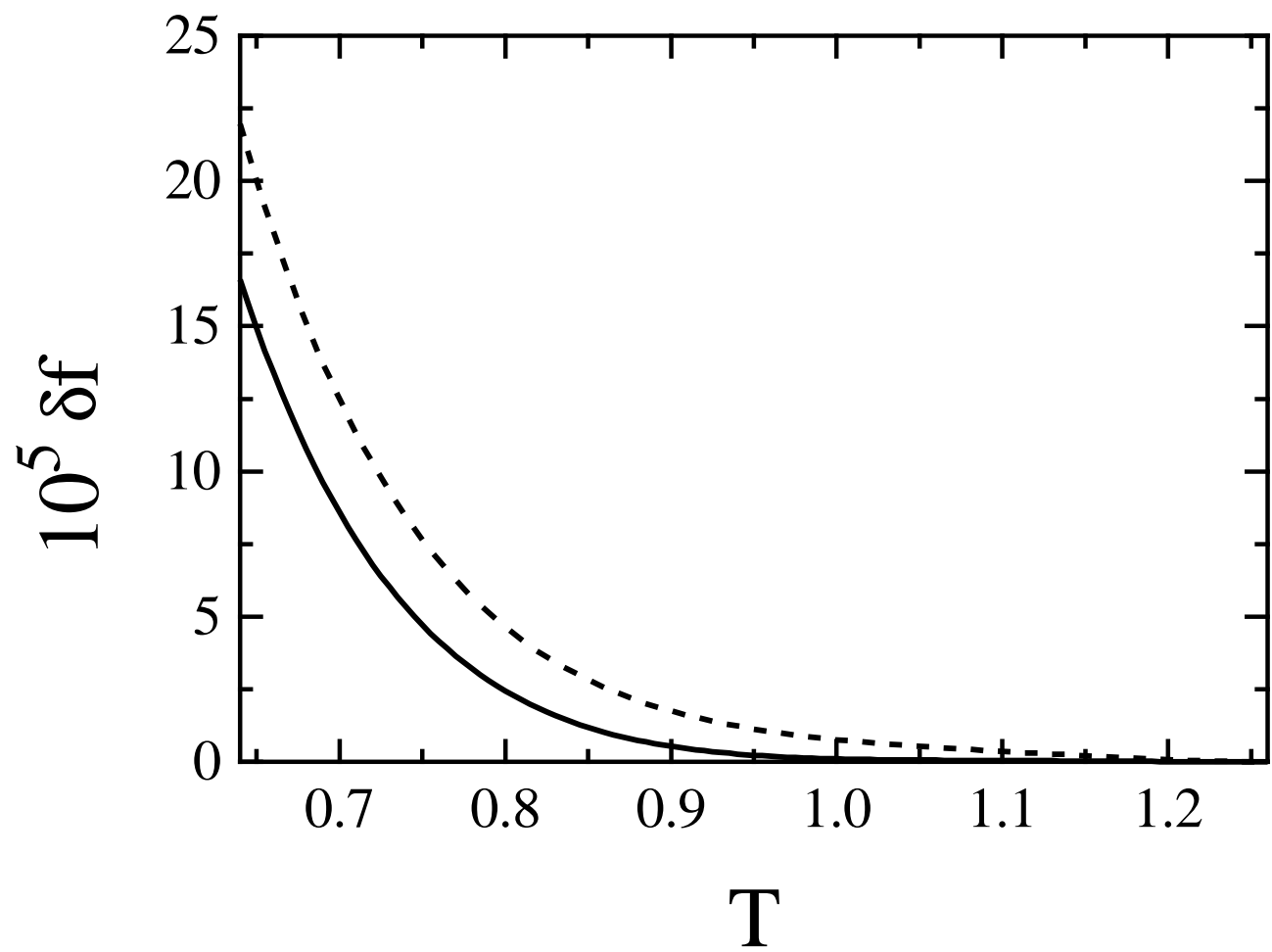
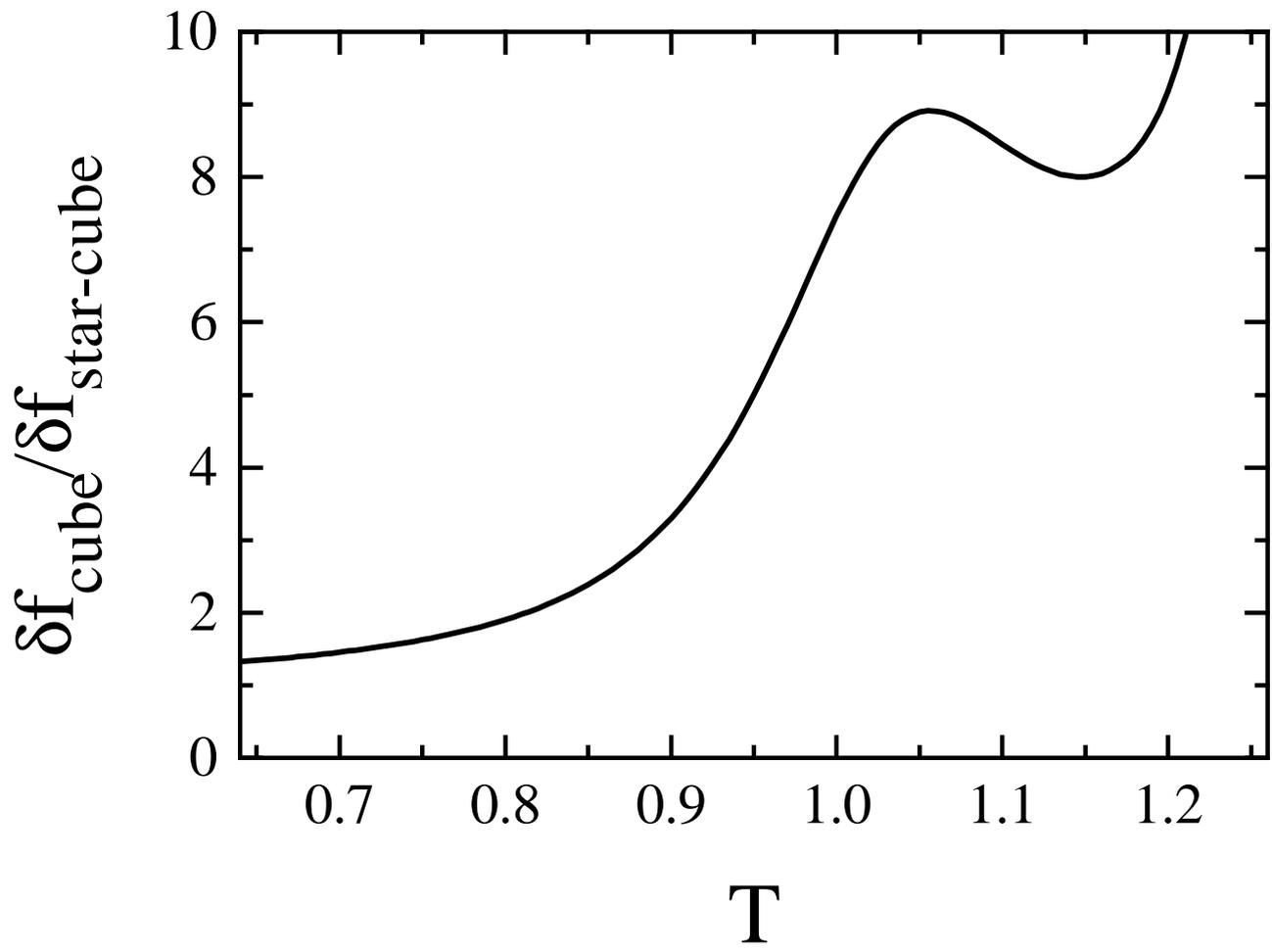
\end{document}